\documentclass{mem}
\usepackage{natbib}
\usepackage{txfonts}
\usepackage{balance}
\usepackage{graphicx}
\usepackage{flushend}
\usepackage[a4paper,breaklinks,pdftex]{hyperref}
\idline{75}{282}
\begin{document}
\def\teff{$T\rm_{eff }$}
\def\kms{$\mathrm {km s}^{-1}$}

\title{
Performant feature extraction for photometric time series
}

   \subtitle{}

\author{
A.\, Lavrukhina\inst{1} 
\and K.\, Malanchev\inst{2,3}
          }

\institute{
Lomonosov Moscow State University, Faculty of Space Research, Russia
\and
Lomonosov Moscow State University, Sternberg Astronomical Institute, Russia
\and
University of Illinois at Urbana-Champaign, Department of Astronomy, USA
\\
\email{lavrukhina.ad@gmail.com}
}

\authorrunning{Lavrukhina}

\titlerunning{Performant feature extraction for photometric time series}

\date{Received: Day Month Year; Accepted: Day Month Year}

\abstract{
Astronomy is entering the era of large surveys of the variable sky such as Zwicky Transient Facility (ZTF) and the forthcoming Legacy Survey of Space and Time (LSST) which are intended to produce up to a million alerts per night.
Such an amount of photometric data requires efficient light-curve pre-processing algorithms for the purposes of subsequent data quality cuts, classification, and characterization analysis.
In this work, we present the new library "light-curve" for Python and Rust, which is intended for feature extraction from light curves of variable astronomical sources.
The library is suitable for machine learning classification problems: it provides a fast implementation of feature extractors, which outperforms other public available codes, and consists of dozens features describing shape, magnitude distribution, and periodic properties of light curves.
It includes not only features which had been shown to provide a high performance in classification tasks, but also new features we developed to improve classification quality of selected types of objects.
The "light-curve" library is currently used by the ANTARES, AMPEL, and Fink broker systems for analyzing the ZTF alert stream, and has been selected for use with the LSST.
}
\maketitle{}

\begin{figure*}[t!]
\resizebox{\hsize}{!}{\includegraphics[clip=true]{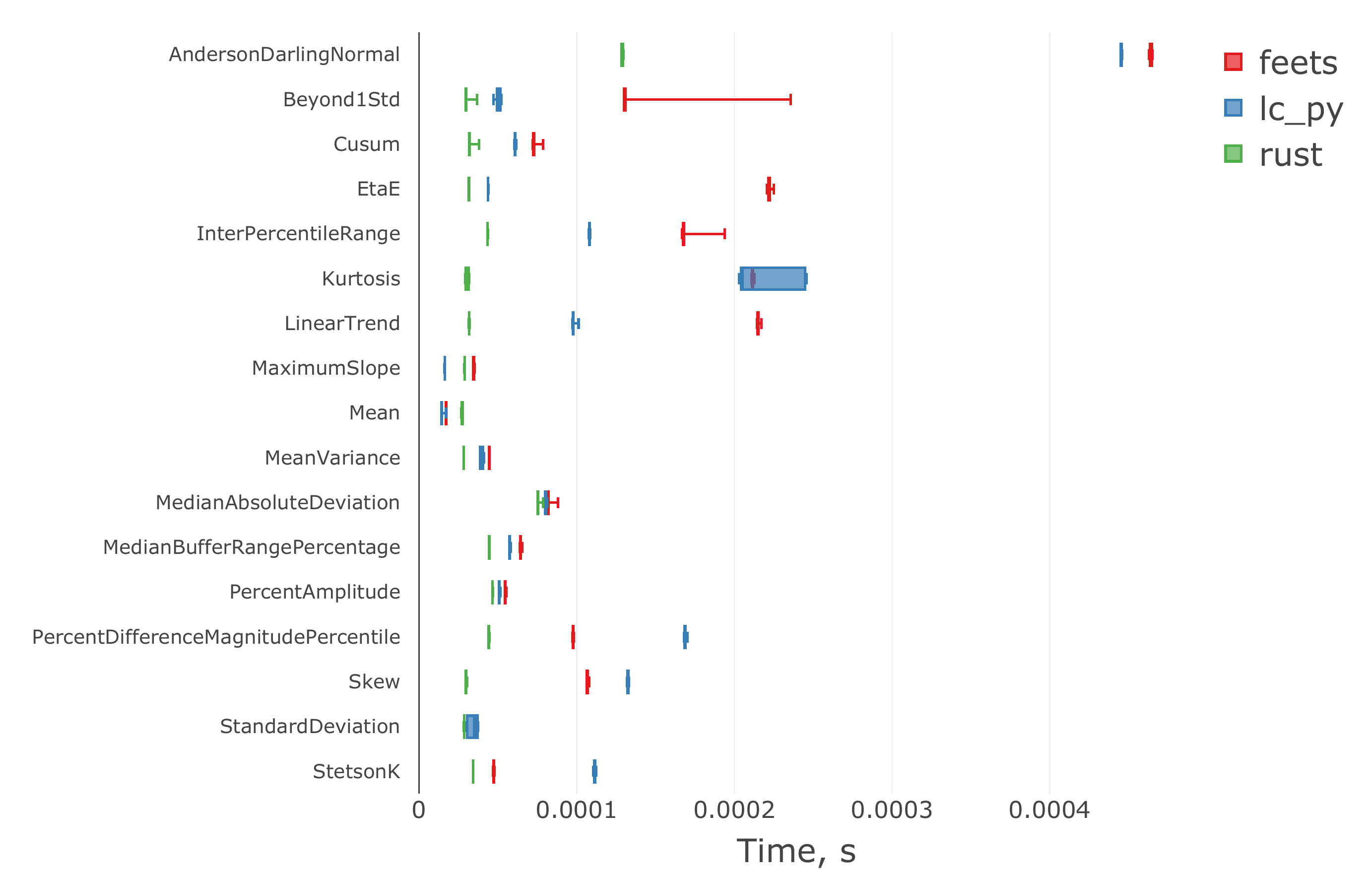}}
\caption{\footnotesize Benchmark boxplots for the features which are implemented in all discussed packages: "feets", "Rust" implementation of "light-curve" package, and its Python sub-package. The x-axis is the time in seconds of feature extraction from the light curve consisting of 1,000 data points. "feets" library performance is labeled as "feets", Python version of "light-curve" as "lc\_py" and the Rust one as "rust". The median value, the maximum, the minimum, the first and the third quartiles were counted.}
\label{benchmark}
\end{figure*}

\begin{figure*}[t!]
\resizebox{\hsize}{!}{\includegraphics[clip=true]{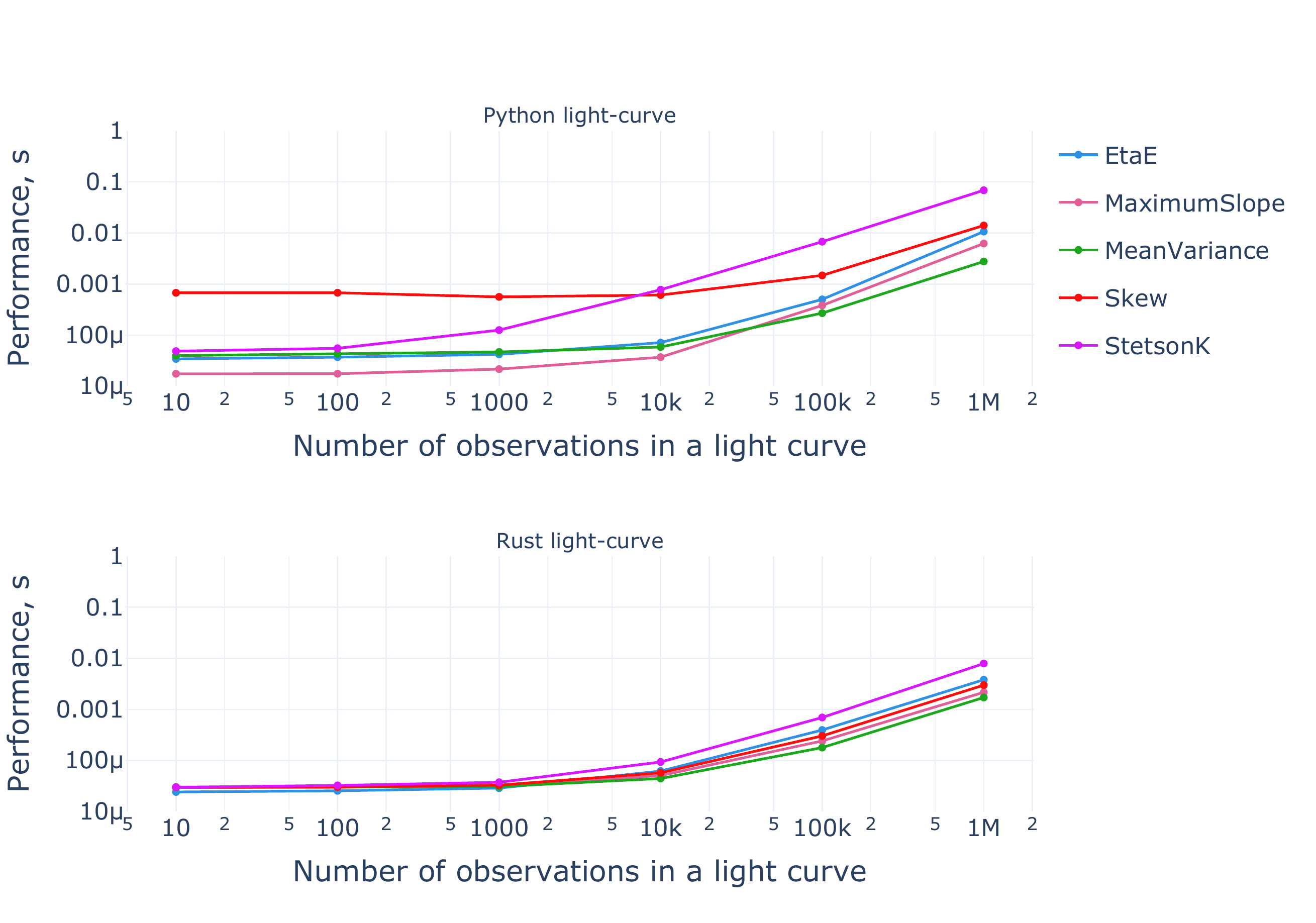}}
\caption{Benchmark results of several features for both the pure-Python and Rust implementations of the "light-curve" package, as a function of the number of observations in a light curve. Both the x-axis and y-axis are on a logarithmic scale.}
\label{benchmark_nobs}
\end{figure*}

\begin{figure*}[t!]
\resizebox{\hsize}{!}{\includegraphics[clip=true]{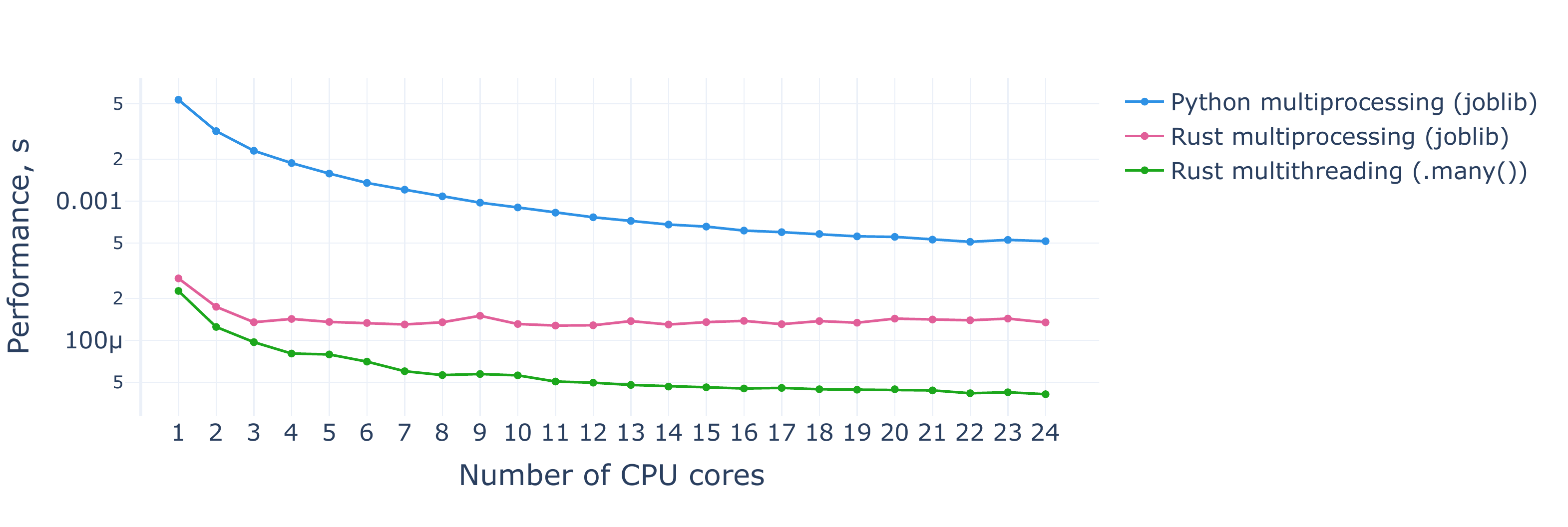}}
\caption{Processing time per a single light curve for extraction of features subset presented in Fig.~\ref{benchmark} versus the number of CPU cores used. The blue curve represents pure-Python implementation and "Joblib"'s multiprocessing strategy, the red curve represents the same strategy for the Rust implementation, and the green curve represents the built-in multithreading implementation with the Rust sub-package. The dataset consists of 10,000 light curves with 1,000 observations in each.}
\label{benchmark_multi}
\end{figure*}

\section{Introduction}

Modern astronomical surveys contain a large amount of data about billions of astronomical sources.
For example, the Zwicky Transient Facility (ZTF)~\citep{ztf} obtains about 1 million alerts per night, which total volume is over 70 GB~\citep{ztf_system}.
For this reason, there is an urgent need for productive and convenient in use methods of data processing. 
In this work, we present a new Rust/Python package for feature extraction from light curves of astronomical sources named "light-curve"~\citep{2021ascl.soft07001M} and compare its performance with another tool -- "feets" library~\citep{cabral2018fats}, which is written entirely in Python and based on "Numpy"~\citep{numpy}, "SciPy"~\citep{scipy}
and "statsmodels"~\citep{statsmodels}.

\section{light-curve Python package}
The "light-curve" project aims to bring high-performance processing of irregularly sampled time-series data in Rust and Python.
Currently, it consists of several Rust packages including "light-curve-feature" crate for feature extraction, which was developed as a part of SNAD anomaly-detection pipelines~\citep{2021ascl.soft07001M, Aleo, Masha}.
While this package has proven itself as a fast and thread-parallelizable solution, we wrapped it into "light-curve" Python package.
However, this package consists of two parts: the wrapper sub-package and a pure Python sub-package implemented with "Numpy" and "SciPy". 
The wrapper sub-package offers high-performance Rust implementation of features with memory-safe Python interoperability, while the second sub-package is utilized for the development of new experimental features.
The availability of two separate implementations of the same feature extractors enables us to validate the correctness of both approaches and measure the performance boost provided by the Rust sub-package.

\section{Benchmark}
For the first benchmark, we used the subset of features, which are implemented in both "light-curve" and "feets" packages. 
Each feature was extracted from a randomly generated light curve consisting of 1,000 observations. 
The benchmark results are shown in Fig.~\ref{benchmark}.
See the feature descriptions in~\cite{2021ascl.soft07001M} or \url{https://docs.rs/light-curve-feature/0.5.2/light_curve_feature/features/index.html}.

The second benchmark, shown in Fig.~\ref{benchmark_nobs}, examines the performance of several feature extractors, including both pure Python and Rust implementations, as the number of observations in a light curve increases.
A linear relationship between the number of observations and the performance of the feature extractors is observed, with a noticeable impact on performance when the number of observations exceeds 10,000.
It is likely due to overhead such as checking if the time array is sorted and ensuring the magnitude array does not contain inappropriate values.

The last benchmark (see Fig.~\ref{benchmark_multi}) is intended to present the dependency of consumed time from the used number of CPU cores for the feature subset, presented in Fig.~\ref{benchmark}. 
The features are extracted from a dataset of 10,000 light curves, each consisting of 1,000 observations.
The multiprocessing strategies include "Joblib"\citep{joblib} for the pure Python sub-package and Rust bindings, as well as the built-in multithreading facility provided by the \texttt{many()} method in the Rust bindings.
In situations where many features need to be extracted from a large dataset of light curves, the Rust multithreading approach shows the best performance, although performance is still impacted by overhead from the multithreading, such as data serialization and deserialization, and message passing between parent and child threads.

The hardware setup for the benchmarking was a dual-CPU Intel Xeon Gold 5118 server.

\section{Conclusions}
The results of the test show that the Rust version of "light-curve" package outperforms other implementations by a factor 1.5-50. 
This superiority is due to Python's dynamic typing and interpreted language nature.
Such performance enables the extraction of a large set of "cheap" features in just a few milliseconds per CPU core for 1,000 observations.
The library also features fast implementations of the periodogram~\citep{Lomb, Scargle} and parametric fits~\citep{Bazin, Villar}.
The time to extract the feature set, including the periodogram, Bazin and Villar fits, is $\approx 25$ ms $\times$ CPU for six ugrizy LSST 3-year light curves.
It allows the processing of all LSST alerts in real-time using a single computational node.

\begin{acknowledgements}
The reported study was funded by RFBR, Russia and CNRS, France according to the research project No. 21-52-15024. 
\end{acknowledgements}

\bibliographystyle{aa}
\bibliography{lavrukhina}

\end{document}